\documentclass[aps,pre,twocolumn,groupedaddress,showpacs]{revtex4-1}
\usepackage[dvips]{graphics}
\usepackage{vmargin}
\setpapersize{USletter}
\setmarginsrb{0.8in}{.6in}{0.8in}{0.8in}{0in}{0.2in}{0in}{0in}
\begin{document}

\title{Localization phase diagram of two-dimensional quantum percolation}
\author{Brianna S. Dillon}\email{dillonb@purdue.edu}
\author{Hisao Nakanishi}\email{hisao@purdue.edu}
\affiliation{Department of Physics, Purdue University, West Lafayette, IN 47907}
\date{\today}

\begin{abstract}
We examine quantum percolation on a square lattice with random dilution up to $q=38\%$ 
and energy $0.001 \le E \le 1.6$ (measured in units of the hopping matrix element), 
using numerical calculations of the transmission coefficient at a much larger scale than 
previously. Our results confirm the previous finding that the two dimensional quantum 
percolation model exhibits localization-delocalization transitions, where the localized 
region splits into an exponentially localized region and a power-law localization region.  
We determine a fuller phase diagram confirming all three regions for energies as low as 
$E=0.1$, and the delocalized and exponentially localized regions for energies down to 
$E=0.001$. We also examine the scaling behavior of the residual transmission coefficient 
in the delocalized region, the power law exponent in the power-law localized region, and 
the localization length in the exponentially localized region. Our results suggest that
the residual transmission at the delocalized to power-law localized phase boundary may
be discontinuous, and that the localization length is likely not to diverge with a 
power-law at the exponentially localized to power-law localized phase boundary.
However, further work is needed to definitively assess the characters of the two phase
transitions as well as the nature of the intermediate power-law regime.
\end{abstract}


\maketitle

\section{Introduction}
Quantum percolation is one of the common models of quantum transport through disordered media.
While transport in classical percolation hinges on whether or not geometrically spanning
or percolating path exists through the medium, quantum percolation refers to a problem of
whether or not the quantum mechanical wave function in a randomly diluted medium (i.e.,
a medium subject to percolation disorder) has a localized character. 
For example, even in a completely ordered system (fully $occupied$ in percolation 
terminology), a quantum mechanical particle may
exhibit zero or very low transmission depending on the details such as the energy of the particle 
or the boundary condition because of the quantum interference effects \cite{cuansing:04}.

It also differs in several important respects from the Anderson model, another well known 
model of localization in a disordered, quantum mechanical system. For example,
whereas, in the Anderson model, the disorder is a continuous distribution of finite width
in the diagonal terms of the tight-binding Hamiltonian, in the quantum percolation model,
the disorder is expressible as a binary distribution of zero or a non-zero finite value in
the off-diagonal terms of the quantum percolation Hamiltonian:

\begin{equation}
H = \sum_{<ij>} V_{ij}|i\rangle \langle j| + h.c
\label{eq1}
\end{equation}

where $|i\rangle$ and $|j\rangle$ are tight binding basis functions at sites $i$ and $j$,
respectively, and $V_{ij}$ is a binary random variable representing the hopping matrix element 
between sites $i$ and $j$, which is equal to a finite constant $V_0$ if $i$ and $j$ are both
$available$ and are nearest neighbors and to zero otherwise. In this work, we use
$V_0 = 1$ as the nominal standard value, which sets the overall scale of energy.
The $availability$ of the sites is subject to the standard (site) percolation disorder, i.e., 
independently from site to site with probability $1-q$, where $q$ is the dilution fraction. 
Because of the difference in the nature of disorder,
the commonly accepted notion that all states are exponentially localized in thermodynamic 
limit in $d=2$ dimensions \cite{abrahams:79} for the Anderson model does not necessarily 
hold true for quantum percolation. 
Indeed, there is a continuing controversy as to the possible phases 
of quantum percolation (for a review, see, e.g., Moorkerjee et al \cite{mookerjee:09}, 
Nakanishi and Islam \cite{nakanishi:09}, Schubert and Fehske \cite{schubert:09} and references therein).
It is to be noted that, even with the Anderson model itself, the presence of non-localized 
or at least only weakly localized states have been seen at or near the band center \cite{eilmes:01}, 
in contradiction to the original one-parameter scaling theory \cite{abrahams:79}.
More recent relevant works on quantum percolation include Ref. \cite{gong:09} and \cite{schubert:08}.

\begin{figure}[t]
{\resizebox{1.7in}{2.7in}{\includegraphics{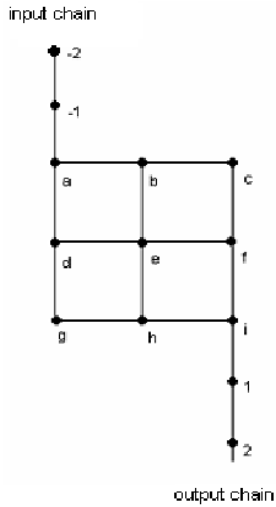}}}\\
\caption{Example of a small system of $3 \times 3$ square lattice cluster with a point-to-point 
connection. The letters label the lattice points of the cluster part of the Hamiltonian, 
while numbers label those of the leads. The same sequence of labeling is used for all sizes
 of the clusters in this work.}
\label{sq_lat}
\end{figure}

This work follows the approach of Islam and Nakanishi \cite{islam:08}, which set up the
problem following the method introduced by Daboul et al \cite{daboul:00} but used a
direct calculation of the transmission instead of a series expansion. However, we do so in a much larger 
scale and over significantly wider ranges of the parameters. This allows us to map out large 
parts of the localization phase diagram of 2D quantum percolation as well as to study
more closely the viability and characteristics of the power-law localized and delocalized 
states suggested there. Thus, we have realized this model on the square lattice of varying sizes
and, using the method of Daboul et al \cite{daboul:00}, calculated the transmission
coefficient from a one-dimensional lead attached to a corner of a square cluster to another
one-dimensional lead attached to the diagonally opposite corner. This is done numerically
but essentially exactly for each realization of the disordered configuration, and final 
data points are obtained by averaging over many such realizations, typically several hundred 
to 1000 for each dilution $q$, energy $E$, and lattice size $L \times L$ that sits between 
the two leads. The geometry of the system is shown in Fig.~\ref{sq_lat}.

The wave function of the entire cluster-lead system can be calculated by solving the time
independent Schr\"{o}dinger equation:

\begin{equation}
\begin{array}{l}
H\psi = E\psi  \\
\mbox{where}, \psi = \left[ \begin{array}{c} \vec{\psi}_{in} \\ 
                     \vec{\psi}_{cluster} \\ \vec{\psi}_{out} \end{array}
               \right]
\end{array}
\label{eq2}
\end{equation}
and $\vec{\psi}_{in} = \left\{\psi_{-(n+1)}\right\}$ and 
$\vec{\psi}_{out} = \left\{\psi_{+(n+1)}\right\}$,
$n = 0,1,2 \ldots$, are the input and output chain part of the wave function respectively.

Daboul et al \cite{daboul:00} proposed to reduce this infinite size problem into a finite
one that involves only the middle cluster and the two sites on the leads that connect to it.
In this approach, an ansatz is used which assumes that the input and output parts of
the wave function are plane waves:

\begin{equation}
\begin{array}{l}
\psi_{in} {\rightarrow} \psi_{-(n+1)} = e^{-inq} + re^{inq} \\
\psi_{out} {\rightarrow} \psi_{+(n+1)} = te^{inq}
\end{array}
\label{eq3}
\end{equation}

where {\it r} is the amplitude of reflected wave, {\it t} is the amplitude of the transmitted wave.
This ansatz is consistent with the Schr\"{o}dinger equation only for the wave vector $q$ that is
related to the energy {\it E} by

\begin{equation}
E = e^{-iq} + e^{iq}
\label{eq4}
\end{equation}

Using this ansatz along with the energy restriction Eq.~(\ref{eq4}), the matrix equation for a
$N \times N$ cluster connected to semi-infinite chains (Fig.~\ref{sq_lat} for an example 
where $N=3$) reduces to an $(N^{2}+2) \times (N^{2}+2)$ matrix equation of the following form (for
details see Daboul et al.\cite{daboul:00}):

\begin{equation}
\left[ \begin{array}{ccc}
         -E + e^{iq}  & \vec{c_1}^t  &   0 \\
               \vec{c_1} & \begin{array}[t]{c}
                             {\bf A}
                         \end{array}              &    \vec{c_2} \\
               0     & \vec{c_2}^t  &  -E + e^{iq}
          \end{array} \right]
          \left[ \begin{array}{c}
                    1 + r \\ \vec{\psi}_{clust} \\ t
                    \end{array} \right]
         = \left[ \begin{array}{c}
                  e^{iq} - e^{-iq} \\ \vec{0} \\ 0
                  \end{array} \right]
\label{eq5}
\end{equation}

where {\bf A} is a $N^2 \times N^2$ matrix representing the connectivity of the cluster
(also with $-E$ as its diagonal elements), $\vec{c_i}$ is the $N^2$-component vector 
representing the coupling of the leads to the corner sites, and $\vec{\psi}_{clust}$ and 
$\vec{0}$ are also $N^2$-component vectors, the former representing the wave function 
solutions (e.g., $\psi_a$ through $\psi_i$ for the cluster in Fig.\ref{sq_lat}). 
With $V_0 = 1$, the cluster connectivity in {\bf A} is represented
with 1 in positions $A_{ij}$ and $A_{ji}$ if sites $i$ and $j$ are connected, otherwise
0.

Eq.~(\ref{eq5}) is the exact expression for a 2D system connected to semi-infinite chains 
with continuous eigenvalues ranging between -2 and +2. The spectrum is continuous because 
it is still effectively infinite and it is non-degenerate except for the reversal of left 
and right. The transmission ($T$) and the reflection ($R$) coefficients are obtained by 
$T = |t|^{2}$ and $R = |r|^{2}$. Since the system is still effectively one-dimensional,
it only allows us to study a subset of the two-dimensional part of transmission
that occurs within the 2D cluster portion. However, it still provides us with a large window
to observe the localization nature of the wave function and transport.

The remainder of this paper is organized as follows: In Section 2, the overall phase
diagram obtained by this work is presented, together with some details of 
the calculations, the extent of the parameter space is probed, and the methods by which
we determine the phases are described. In Section 3, secondary fits are performed to
analyze the behavior of the localization length, residual transmission coefficient, and
power-law exponents, where applicable. Finally, Section 4 presents our summary and discussions.

\section{Calculation of the Transmission Coefficient $T$ and the Phase Diagram} 

\begin{figure*}[htb]
{\resizebox{6.7in}{2.5in}{\includegraphics{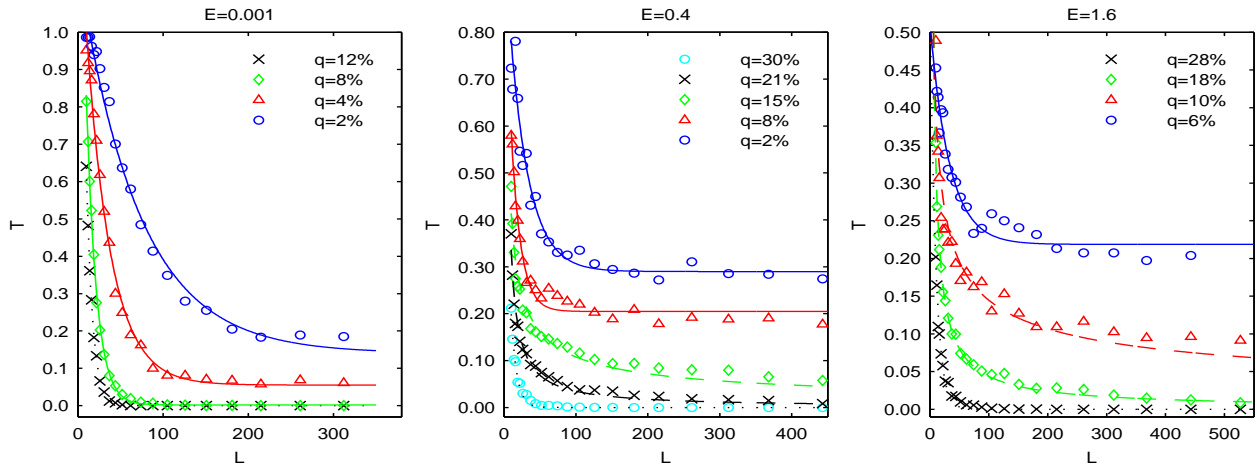}}}\\
\caption{(color online) Representative transmission curves in linear scale at 3 energies and various dilutions q. For each energy and dilution, the type of fit determined to be best is shown. An exponential with offset fit (solid lines) denotes a delocalized state, power-law fit (dashed lines) denotes power-law localization, and pure exponential fit (dotted lines) denotes exponential localization.}
\label{curve_fits}
\end{figure*}

\begin{figure*}[htb]
{\resizebox{6.7in}{2.5in}{\includegraphics{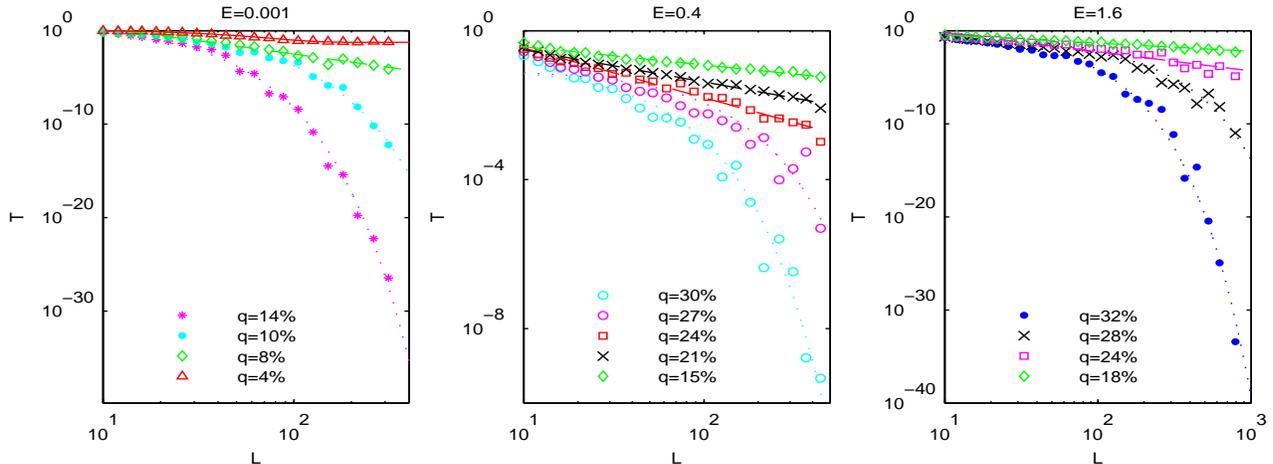}}}\\
\caption{(color online) Representative transmission curves in log-log scale at 3 energies and various dilutions $q$. The type of fit judged as best is shown using the same line types as in Fig.~\ref{curve_fits}.}
\label{log-log}
\end{figure*}

The transmission coefficient $T$ was calculated at 6 different energies in the range
$0.001 \leq E \leq 1.6$ on 23 to 27 different sizes $L \times L$ depending on $E$, where 
$10 \leq L \leq 891$. The maximum dilution range was $0.02 \leq q \leq 0.38$, with increments of 0.01 
to 0.05 depending on the simulation batch. Obtaining sufficient data for the phase diagram required 
multiple batches  in which the average transmission coefficient $T$ was calculated over successively 
smaller ranges of dilution $q$ at each energy and lattice size $L$ in order to narrow the location of the 
transition point, as determined using the procedure summarized below.

For each energy and dilution studied, the transmission coefficient $T$ is fitted numerically
against lattice size $L$, the best fitting form indicating the localization state of the 
quantum particle. In general, an exponential with offset ($T = a\exp (-b L)+c$) was the best fit for the lowest dilutions (beginning at $q=0.02$), suggesting the state is delocalized.  As dilution is 
increased the transmission curve becomes best characterized by a power-law ($T = a L^{-b}$), indicating the particle may be power-law localized, and at the highest dilutions it follows a pure 
exponential ($T = a\exp(- L/l)$), meaning the state is exponentially localized. 
 
At lower dilutions the transmission curve was fit in linear scale and determining best fit 
was straightforward as seen in Fig.~\ref{curve_fits}.  At higher dilutions the distinction between a 
power-law fit and an exponential fit are generally not conclusive, either visually or analytically by looking 
at the coefficient of regression $R_\sigma^2$ of the two fits. In those cases, we instead fit the 
transmission curve in a log-log scale, since there is a distinct visual difference between power-law and 
exponential behavior in that scale. In most cases the switch from power-law behavior to pure 
exponential behavior as dilution increases was visually apparent; the best fit was determined by using 
regression coefficients for completeness sake even in those cases, but often visual  examination was 
sufficient. See Fig.~\ref{log-log} for example.

Also important to note is that in log-log scale, the small $L$ portion of an exponential transmission curve 
looks linear, which necessitated including larger lattice sizes for higher dilutions to increase the certainty 
that a dilution with a power-law fit really did belong in the power-law localized region and not the 
exponentially localized region. There is, of course, a possibility that if still larger lattice sizes were 
included in the transmission curve, the dilutions which were thought to belong to the power-law localized 
region would turn out to actually belong in the exponentially localized regime. 

The transition point between regions was estimated by averaging the upper bound for one state 
and the lower bound for the other. For example, at $E=1.6$ the particle is conclusively 
delocalized for dilutions up to 8\% but is conclusively power-law localized beginning at 
dilutions of 9\%, thus the transition point is estimated to be around 8.5\%. For some batches 
narrowing the step size did not narrow down the location of the transition point; the state 
would be conclusively power-law, for instance, up to some dilution $q_p$,  then switch back and forth 
between power-law and exponential for dilutions $q_p < q < q_e$, then be conclusively exponential for 
all $q > q_e$. In that case the transition point was still estimated as the average of $q_p$ and $q_e$, 
but associated with a larger uncertainty. Successive batches at the same energy which showed slightly 
different bounds for the three regions were treated in a similar manner when estimating the overall 
transition point from those batches used in the proposed phase diagram.  

\begin{figure}[t]
{\resizebox{3in}{2in}{\includegraphics{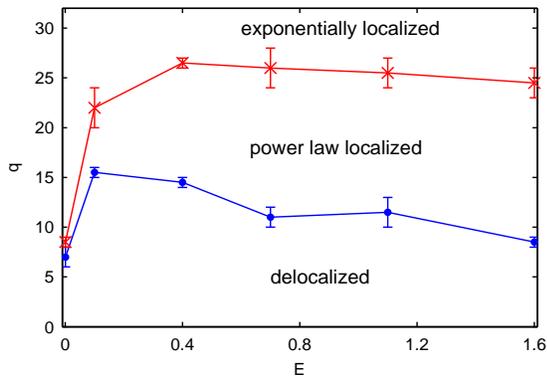}}}\\
\caption{(color online) Phase diagram for the 2D quantum percolation model with site-based dilution
obtained from this work. The presence of the delocalization region at all energies suggests 
that conduction is possible in this model. The lines between points are included merely to 
guide the eye.}
\label{phase_diagram}
\end{figure}

The phase diagram obtained by these calculations is shown in Fig.~\ref{phase_diagram}; the 
bipartite nature of the lattice means that the diagram is symmetric about $E=0$. The results at 
$E=0.001$ and $E=1.6$ agree with our previous work (see, e.g., Islam and Nakanishi \cite{islam:08}) 
both in the presence of three regions and in the location of the transitions between them. 
The three regions grow closer together as the energy decreases to the band center, with the 
power-law region appearing to vanish around $E=0.001$. If there is a power-law region at this 
energy, it is very narrow and was not visible in our calculations. However, the delocalized state 
appears to be always present, suggesting that conduction is possible in this model. 

It is also noted that the minimum $q$ for which the exponential localization is observed
is rather flat for $E \geq 0.4$, below which it sharply decreases toward the band center ($E=0$).
These observations are in good agreement with the findings of Gong and Tong \cite{gong:09} who
used a method centered on von Neumann entropy and also with an earlier work of Koslowski and
Niessen \cite{koslowski:90} who used the sensitivity of eigenvalues to the boundary conditions.

\section{Further Analysis}

In order to assess how faithfully our phase diagram represents system behavior in the 
thermodynamic limit, especially regarding the existence of the delocalized state, we examined 
how the relevant calculated parameters depend on the fitting region. For the delocalized region, 
we study the offset term $c$ that appears in the exponential plus offset fits 
$T \sim a\exp (-b L)+c$, where a non-zero value indicates the residual transmission as $L \rightarrow 
\infty$. For the power-law localized region, we investigate the power-law exponent $b$ in $T \sim a 
L^{-b}$. For the exponentially localized region, we focus on the parameter $l^{-1}$ in the fit
$T \sim a\exp (- L/l)$, which is the inverse of the localization length $l$.

For each transmission curve (i.e., for a given $E$ and $q$ and varying lattice size $L$), we fit
successively larger numbers of points, at first using only the smallest values of $L$  
(first 10 for inverse localization length, first 15 for residual transmission and power-law exponent), 
and then successively adding points corresponding to larger $L$ until the entire curve has been 
used. In all cases the successive fits were done in a linear scale regardless of the dilution, 
since this yielded the most precise parameter estimates \cite{comment}. Relevant parameters ($c$, $b$, 
or $l^{-1}$) are then plotted against $L_{max}$, the maximum value of $L$ included in the fit, to 
determine whether the parameters reach a stable value ($c_{\infty}$, $b_{\infty}$, or 
$l_{\infty}^{-1}$) as more points were included and $L_{max} \rightarrow \infty$.

\begin{figure}[!t]
{\resizebox{3in}{2.5in}{\includegraphics{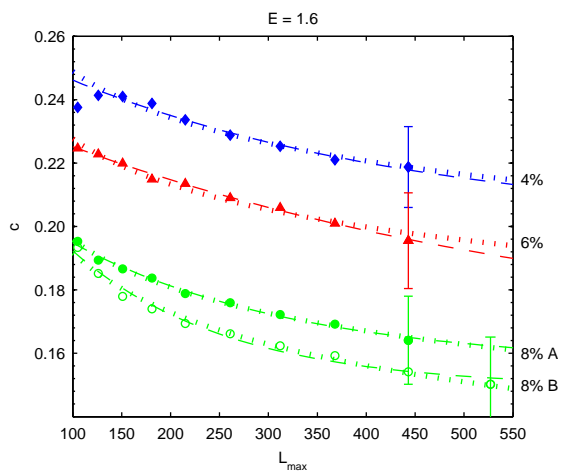}}}\\
\caption{(color online) For the delocalized region: The estimated offset parameter $c$ is plotted vs. $L_{max}$, the maximum lattice size included in the fit to estimate $c$, for dilutions within 
the delocalized region at energy $E=1.6$. In some cases, calculations were based on different 
simulation batches at the same dilution. Points were further fit with an exponential 
with offset (dashed line) and with a power-law (dotted line) to determine
a limiting estimate $c_{\infty}$.}
\label{c_vs_Lmax}
\end{figure}

For the delocalized region, $c$ does indeed appear to stabilize to a nonzero value. This is
illustrated with a typical example from $E=1.6$ in Fig.~\ref{c_vs_Lmax}. We further fitted the points in 
diagrams such as Fig.~\ref{c_vs_Lmax} to quantify this stabilization with both another exponential with 
an offset and a power-law. Both yielded excellent fits, with  $R_\sigma^2 \approx 0.98-0.99$ in all but 
a few cases. Although the difference between the quality  of the two fit types was not significant, in the 
majority of the cases the exponential with offset was still the better one. With this type of a fit, in all 
cases the limiting value $c_{\infty}$ excluded zero taking into account the fitting uncertainties. This 
gives us some confidence that the non-zero transmission persists in the thermodynamic limit.

\begin{figure}[!b]
{\resizebox{3in}{3.5in}{\includegraphics{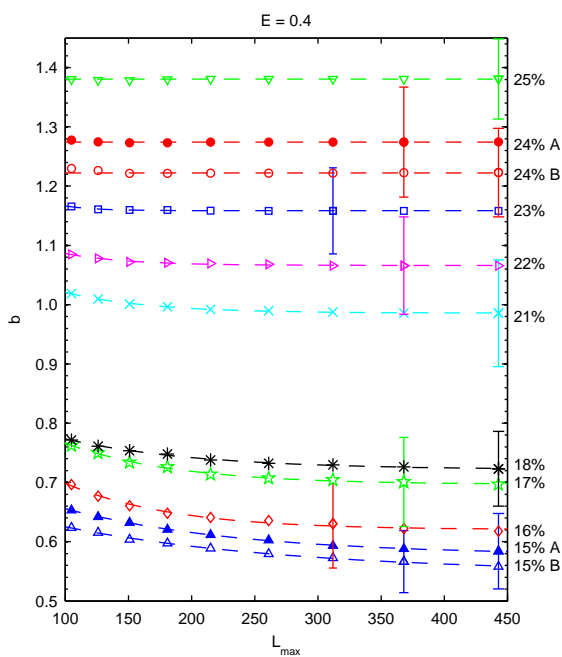}}}\\
\caption{(color online) For the power-law localized region: The power law exponent $b$ is plotted vs. $L_{max}$ similarly to Fig.~\ref{c_vs_Lmax}, for dilutions within the power-law region at energy $E=0.4$. In some cases, different batches at the same dilution were used. To determine a limiting estimate $b_{\infty}$ as $L_{max} \rightarrow \infty$, these points were either further fit with an exponential with offset or averaged over all but the first few points before fluctuations dissipated, 
as shown by the dashed lines.}
\label{power_law_fits}
\end{figure}

In the power-law localized region, the exponent of the power-law $b$  was analyzed in the same 
manner as the residual transmission $c$ for the delocalized region. The estimate of $b$ stabilized,
in most cases very quickly, as in the example shown in Fig.~\ref{power_law_fits}. For some dilutions, 
the estimated $b$ vs $L_{max}$ data were further fitted by another exponential with offset to 
determine the limiting value $b_{\infty}$, as with an example of $q \le 23\%$ in 
Fig.~\ref{power_law_fits}. For other dilutions, the fitted $b$ vs $L_{max}$ data were so flat as to 
render further fitting impossible (such as the case of $q=24\%$ and $25\%$ in 
Fig.~\ref{power_law_fits}), in which case $b_{\infty}$ was determined by simply averaging $b$ over all 
but the first few points, for which the uncertainty was still decreasing. The resulting estimates of 
$b_{\infty}$ increase as the dilution is increased from the delocalized/power-law boundary to the power-
law/exponentially localized boundary. 

\begin{figure}[!b]
{\resizebox{3in}{4in}{\includegraphics{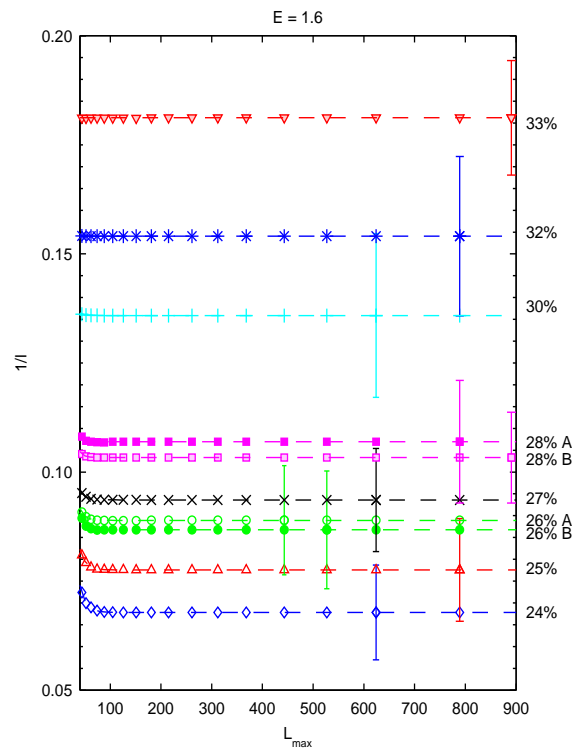}}}\\
\caption{(color online) For the exponentially localized region: The estimated inverse localization length $l^{-1}$ is plotted vs. $L_{max}$ similarly to Fig. 5 and 6, for dilutions within the exponentially localized region at energy $E=1.6$. In some instances different batches were used at the same dilution. To determine a limiting estimate $l_{\infty}^{-1}$ as  $L_{max} \rightarrow \infty$ these points were either further fit with an exponential with offset or averaged over all but the first few points before fluctuations dissipated, as shown by the dashed lines.}
\label{exp_fits}
\end{figure}

For the exponentially localized region, the inverse localization length estimates $l^{-1}$ stabilized 
as $L_{max} \rightarrow \infty$ similarly to the exponent estimates in the power-law localized
region above. An example is shown for $E=1.6$ in Fig.~\ref{exp_fits}. As before, the
estimated $l^{-1}$ vs $L_{max}$ data were further fitted by another exponential with offset to determine the limiting value $l_{\infty}^{-1}$ for some dilutions, as with an example of $q \le 26\%$ shown in Fig.~\ref{exp_fits}. For still larger dilutions, the fitted $l^{-1}$ vs $L_{max}$ data were flat (such as for $q=30\%-$33\% in Fig.~\ref{exp_fits}), in which case $l_{\infty}^{-1}$ was determined by averaging $l^{-1}$ over all but the first few points, for which the uncertainty was still decreasing.

\begin{figure}[t]
{\resizebox{3in}{3in}{\includegraphics{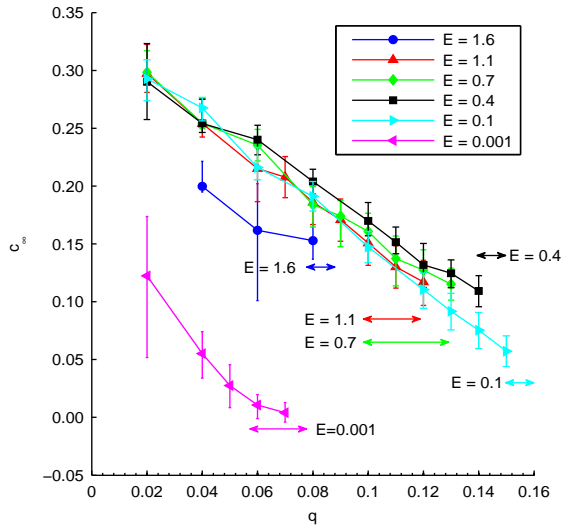}}}\\
\caption{(color online) The stable values $c_{\infty}$ of the delocalization offset term are plotted vs the dilution q of the lattice on which they were calculated for all energies studied. Lines between points are included to guide the eye and are not reflective of any fit. The double arrows denote the transition bounds for each energy as shown in Fig.~\ref{phase_diagram}. }
\label{CvsQ}
\end{figure}

\begin{figure}[htbp]
{\resizebox{3in}{3in}{\includegraphics{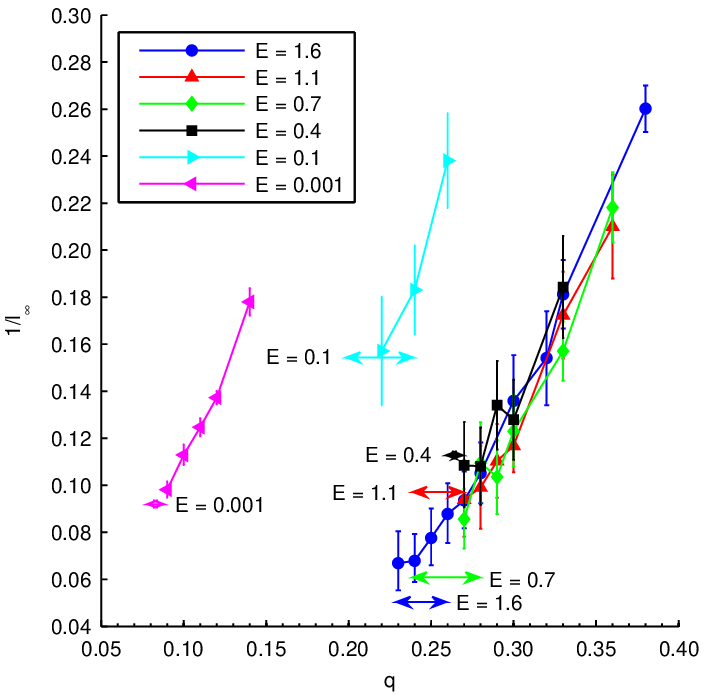}}}\\
\caption{(color online)  The stable values $l_{\infty}^{-1}$ of the inverse localization length are plotted vs the dilution 
$q$ of the lattice on which they were calculated for all energies studied. Lines between points are 
included to guide the eye and are not reflective of any fit. The double arrows denote the transition 
bounds for each energy as shown in Fig.~\ref{phase_diagram}.}
\label{ExpBvsQ}
\end{figure}

\begin{figure}[htbp]
{\resizebox{3in}{3in}{\includegraphics{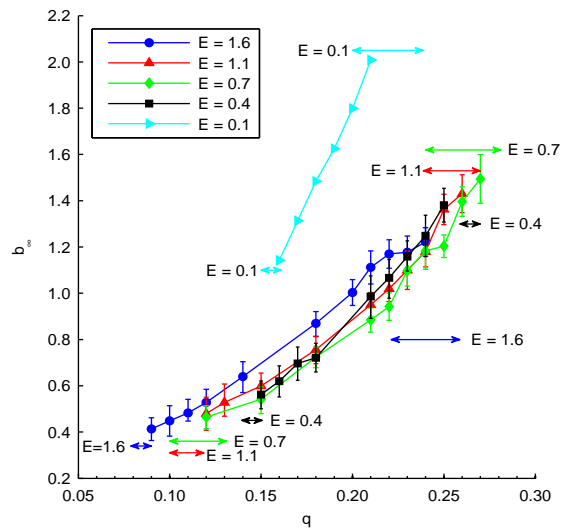}}}\\
\caption{(color online)  The stable values $b_{\infty}$ of the power law exponent are plotted vs the dilution $q$ of the lattice on which they were calculated, for all energies studied which showed a power-law localized region. Lines between points are included to guide the eye and are not reflective of any fit. 
The uncertainty for points on the $E=0.1$ curve are on the order of the size of the points. 
The double arrows denote the transition bounds for each energy as shown in Fig.~\ref{phase_diagram}.}
\label{PwrBvsQ}
\end{figure}

Having obtained the stable values of the relevant parameters for each dilution and energy studied, we 
next plotted the stable values versus dilution $q$ as shown in Fig.~\ref{CvsQ}, Fig.~\ref{ExpBvsQ} 
and Fig.~\ref{PwrBvsQ}. In the delocalized region (Fig.~\ref{CvsQ}), we see that the residual transmission $c_{\infty}$ decreases toward the transition to the power-law localized region, as expected. For the energy $E=0.001$, closest to the band center, $c_{\infty}$ decreases to nearly zero. On the other hand, for energies $E \ge 0.1$, $c_{\infty}$ remains finite for even the dilutions closest to the transition point, suggesting a discontinuity in $c_{\infty}$ at the transition for those energy ranges.

In the exponentially localized region (Fig.~\ref{ExpBvsQ}) the inverse localization length $l^{-1}$
decreases as expected in approaching the power-law region. However, it does not appear to decrease 
to zero, as one would expect for a critical phase transition. Rather, at each energy the inverse 
localization length seems to approach a finite value, possibly suggesting a discontinuity at the 
exponential/power-law localization transition, which may be reminiscent of the results of recent work 
on two dimensional melting by Bernard and Krauth \cite{bernard:11}. However, we cannot rule out
other possibilities such as the transition point simply being overestimated or the $l^{-1}$ having 
a non-power-law behavior that only reveals its tendency toward zero extremely near the transition point.

At first glance it is also curious that the $l^{-1}$ vs $q$ curves for $E \ge 0.4$ all fall nearly on 
top of each other without any rescaling, while those for $E=0.1$ and $E=0.001$ do not. 
However, this peculiarity may only be a reflection of the fact that the energies $E \ge 0.4$ have 
very similar values of the transition points $q_c$ and it is possible that $l^{-1}$ mainly depends
on $q$ and not as much on $E$. A true, two-variable scaling would be needed to see if data collapsing
as in usual critical phenomena occurs and, in that case, it may be possible to also bring in the
outlying data from $E = 0.1$ and $0.01$ to the same curve.

The curves for the estimated power-law exponents $b$ (Fig.~\ref{PwrBvsQ}) for $E \ge 0.4$ likewise all 
nearly overlap with each other. The exponent appears to vary across the power-law region over a wide
range of about $0.2$ to $2.0$, indicating a non-universal behavior in this region, somewhat reminiscent
of below-$T_{KT}$ region of the Kosterlitz-Thouless transition \cite{KT:73}.

\section{Summary and Discussion}

Our current analysis indicates that, at all energies except possibly very near $E = 0$, 
the quantum percolation model in two dimensions possesses delocalized, power-law localized, 
and exponentially localized regions. The power-law region becomes narrower as the energy approaches 
the band center, and possibly disappears there. This is consistent with our own previous more 
limited-scale work \cite{islam:08}, but the current work covers a wider parameter space and 
with larger lattice sizes $L$ and better statistics, allowing systematic extrapolations to 
the thermodynamic limit. Various aspects of our findings are also consistent with other previous
works such as Ref. \cite{gong:09,schubert:08,daboul:00}.

Additionally, the inverse localization length extrapolates to non-zero limits as system size $L \rightarrow \infty$ in the exponentially localized region, and the residual transmission coefficient extrapolates to a non-zero limit as well in the delocalized region, as appropriate. The localization length does not appear to
diverge as the dilution $q$ approaches the phase boundary with the power-law region, and (with the 
exception of $E=0.001$) the transmission coefficient does not seem to decrease all the way to zero at 
the delocalization-localization boundary.

At this stage, we cannot make any firm conclusions regarding the precise nature of the transitions 
which were observed numerically. However, the discontinuous change of transmission $T$ at the
delocalized-power-law transition is reminiscent of the discontinuity in the superfluid density of
two-dimensional superfluid transition or the helicity modulus in 2D XY model at the Kosterlitz-Thouless
transition \cite{KT:73}, which is a continuous transition of rather interesting character.
The seeming discontinuity of the inverse localization length $l^{-1}$ at the other boundary, the power-
law to exponentially localized transition, could be an indication that this transition is of first order,
but it could also be a reflection of a difficulty in estimating the boundary accurately or of a non-power-
law nature of the divergence of the localization length there. Aa an example of a correlation length that
diverges by a non-power law, we cite the exponential behavior ($\xi \sim exp \left[ a \left( \Delta T 
\right)^{-1/2}\right] $) of the Kosterlitz-Thouless problem, approaching the transition from $T > T_{KT}$. 
The fact that many previous works which attempted to estimate this exponent came up with values that 
differ over a wide range (see, e.g., Ref. \cite{daboul:00}) might be suggestive. It is also intriguing 
that Soukoulis and Grest \cite{soukoulis:91}, while concluding that the delocalization is possible only at 
$q = 0$, obtained a numerical result where  $\xi \sim exp \left[ a \left( \frac{p}{1-p} \right)^{1/2}\right] 
$ where $p = 1-q$.

Either way, our results are reminiscent of the two-step transition \cite{nelson:79} of the two-dimensional 
melting with the intermediate hexatic phase. The latter exhibits a power-law translational correlation
in the solid phase, only a power-law orientational correlation in a hexatic phase, and exponentially 
decaying correlations in the liquid phase \cite{bernard:11,nelson:79}. The theory of these transitions
indicate both solid-hexatic and hexatic-liquid transitions to be of continuous, Kosterlitz-Thouless type,
but there are some recent experimental claims that the latter transition is discontinuous \cite{bernard:11}. Since the quantum wave function of our quantum percolation problem does contain continuous two-component
symmetry (as all quantum mechanical problems do), it is not entirely unnatural to consider a possibility of
an analogy to the Kosterlitz-Thouless problem. However, for the two-dimensional melting problem to
exhibit an intermediate hexatic phase, two different types of correlations must become short-range at
different temperatures. The quantum percolation problem does not immediately appear to contain such behavior.
We believe that further work is needed to characterize the observed transitions more accurately. At least
it is clear to us that this problem is far from completely resolved as sometimes claimed.

\section*{Ackowledgements}

We thank Purdue University and its Department of Physics for their generous support including 
computing resources.


\begin{thebibliography}{99}
\bibitem{cuansing:04} E. Cuansing and Hisao Nakanishi, Phys. Rev. E, {\bf 70}, 066142 (2004)
\bibitem{abrahams:79} E. Abrahams, P.W. Anderson, D.C. Licciardello and T.V. Ramakrishnan,
                      Phys. Rev. Lett., {\bf 42}, 673 (1979)
\bibitem{mookerjee:09} A. Mookerjee, T. Saha-Dasgupta, and I. Dasgupta, in {\it Quantum and 
                 Semi-classical Percolation and Breakdown in Disordered Solids}, 
                 edited by A. K. Sen, K. K. Bardhan, and B. K. Chakrabarti, Springer, pages 83-108 (2009)
\bibitem{nakanishi:09} H. Nakanishi and Md F. Islam, in {\it Quantum and Semi-classical
                 Percolation and Breakdown in Disordered Solids}, edited by A. K. Sen,
                 K. K. Bardhan, and B. K. Chakrabarti, Springer, pages 109-133 (2009)
\bibitem{schubert:09} G. Schubert and H. Fehske, in {\it Quantum and 
                 Semi-classical Percolation and Breakdown in Disordered Solids}, 
                 edited by A. K. Sen, K. K. Bardhan, and B. K. Chakrabarti, Springer, pages 163-189 (2009)
\bibitem{eilmes:01} A. Eilmes, R. A. R\"{o}mer, and M. Schreiber, Physica B, {\bf 296}, 46 (2001)
\bibitem{gong:09} L. Gong and P. Tong, Phys. Rev. B {\bf 80}, 174205 (2009)
\bibitem{schubert:08} G. Schubert and H. Fehske, Phys. Rev. B {\bf 77}, 245130 (2008)
\bibitem{islam:08} M.F. Islam and H. Nakanishi, Phys. Rev. E, {\bf 77}, 061109 (2008).
\bibitem{daboul:00} D. Daboul, I. Chang and A. Aharony, Eur. Phys. J, {\bf B 16}, 303 (2000)
\bibitem{comment} While we used a log-log scale for some dilutions to determine the power law/exponentially                   	boundary in the phase diagram, we found that using that scale when fitting $t$ for those dilutions over 		successively larger lattices yeilded inconsistent behavior for the parameter $b$. Since the log-log scale 		tends to amplify the behavior of small $L$, we believe it may not be good for precisely determining 			parameter values, even though it was useful for distinguishing which region of localization the 			transmission curve belongs in.
\bibitem{koslowski:90} Th. Koslowski and W. von Niessen, Phys. Rev. B {\bf 42}, 10342 (1990)
\bibitem{bernard:11} E. P. Bernard and W. Krauth, Phys. Rev. Lett, {\bf 107}, 155704 (2011)
\bibitem{KT:73} J. M. Kosterlitz and D. J. Thouless, J. Phys. C, {\bf 6}, 1181 (1973)
\bibitem{soukoulis:91} C. M. Soukoulis and G. S. Grest, Phys. Rev. B, {\bf 44}, 4685 (1991)
\bibitem{nelson:79} D. R. Nelson and B. I. Halperin, Phys. Rev. B, {\bf 19}, 2457 (1979)

\end{thebibliography}
\end{document}